\documentclass[aps,prd,twocolumn,amssymb,showpacs,superscriptaddress,nofootinbib,longbibliography,floatfix]{revtex4-1}

\usepackage{amsfonts}
\usepackage{latexsym}
\usepackage{yfonts}
\usepackage{amsmath}
\usepackage{amssymb}
\usepackage{amsthm}
\usepackage{mathrsfs}
\usepackage{upgreek}
\usepackage{bm}
\usepackage{dcolumn}
\usepackage{epsfig}
\usepackage{graphicx}
\usepackage{latexsym}
\usepackage{rotating}
\usepackage{color}
\usepackage{soul}
\usepackage[english]{babel}

\usepackage{lmodern}
\usepackage[T1]{fontenc}
\usepackage[utf8]{inputenc}

\usepackage{times}
\usepackage{setspace}

\definecolor{mogreen}{rgb}{0.2,0.7,0.2}

\usepackage{setspace}
\usepackage[hang]{footmisc}
\usepackage{lipsum}

\usepackage{color}
\usepackage[unicode=true,pdfusetitle,
 bookmarks=true,bookmarksnumbered=false,bookmarksopen=false,
 breaklinks=false,pdfborder={0 0 0},backref=false,colorlinks=true]
 {hyperref}
\hypersetup{linkcolor=[rgb]{0.1,0.4,0.1}, linktoc=page}
\hypersetup{citecolor=[rgb]{0.6,0,0.1}}
\hypersetup{urlcolor=[rgb]{0.1,0,0.6}}

\DeclareTextFontCommand{\emph}{\sl}

\begin{document}
%%%%%%%%%%%%%%%%%%%%%%%%%%%%%%%%%%%%%%
\title{Relativistic geoids and time dependence: quasilocal frames versus isochronometric surfaces}
%%%%%%%%%%%%%%%%%%%%%%%%%%%%%%%%%%%%%%

%*************************************
\author{Marius Oltean}
%*************************************
\email[]{oltean@ice.cat}

\affiliation{Institute of Space Sciences (CSIC-IEEC), Campus Universitat Aut\`{o}noma de Barcelona, Carrer de Can Magrans s/n, 08193 Cerdanyola del Vall\`{e}s (Barcelona), Spain}

\affiliation{Departament de F\'isica, Facultat de Ci\`{e}ncies, Universitat Aut\`{o}noma de Barcelona, Edifici C, 08193 Cerdanyola del Vall\`{e}s (Barcelona), Spain}

\affiliation{Observatoire des Sciences de l'Univers en r\'{e}gion Centre, Universit\'{e} d'Orl\'{e}ans, 1A rue de la F\'{e}rollerie, 45071 Orl\'{e}ans, France}

\affiliation{P\^{o}le de Physique, Collegium Sciences et Techniques, Universit\'{e} d'Orl\'{e}ans, Rue de Chartres, 45100  Orl\'{e}ans, France}

\affiliation{Laboratoire de Physique et Chimie de l'Environnement et de l'Espace, Centre National de la Recherche Scientifique, 3A Avenue de la Recherche Scientifique, 45071 Orl\'{e}ans, France}

%*************************************
\author{Richard J. Epp}
%*************************************
\email[]{rjepp@uwaterloo.ca}

\affiliation{Department of Physics and Astronomy, University of Waterloo, 200 University Avenue West, Waterloo, Ontario N2L 3G1, Canada}

%*************************************
\author{Paul L. McGrath}
%*************************************
\email[]{pmcgrath@uwaterloo.ca}

\affiliation{Faculty of Mathematics, University of Waterloo, 200 University Avenue West, Waterloo, Ontario N2L 3G1, Canada}

%*************************************
\author{Robert B. Mann}
%*************************************
\email[]{rbmann@uwaterloo.ca}

\affiliation{Department of Physics and Astronomy, University of Waterloo, 200 University Avenue West, Waterloo, Ontario N2L 3G1, Canada}

\affiliation{Perimeter Institute for Theoretical Physics, 31 Caroline Street North, Waterloo, Ontario N2L 2Y5, Canada}

\date{\today}

%%%%%%%%%%%%%%%%%%%%%%%%%%%%%%%%%%%%%%
\begin{abstract}
%%%%%%%%%%%%%%%%%%%%%%%%%%%%%%%%%%%%%%

In a recent paper \cite{philipp_definition_2017}, a definition of a general-relativistic geoid, restricted to stationary spacetimes, was presented in terms of ``isochronometric surfaces''. In this note, we explicate how this definition is just a special, highly restrictive case of our earlier formulation of the general-relativistic geoid using quasilocal frames, which is valid for generic, non-stationary spacetimes \cite{oltean_geoids_2016}. Moreover, like the isochronometric surface geoid, we show how our ``geoid quasilocal frame'' (GQF) can also be defined, simply and operationally, in terms of redshift measurements. 

\end{abstract}

%  PACS  
% \pacs{04.30.Db, 04.40.Dg, 95.30.Sf, 97.10.Sj}

\maketitle

%\pagebreak{}
%\noindent \begin{center}
%\rule{1\columnwidth}{1pt}
%\vspace{-0.8cm}
%\par\end{center}
%\tableofcontents{}
%\noindent \begin{center}
%\par\end{center}
%\noindent \begin{center}
%\rule{1\columnwidth}{1pt}
%\par\end{center}

%%%%%%%%%%%%%%%%%%%%%%%%%%%%%%%%%%%%%%

\noindent \textit{Introduction and literature overview.} \textemdash{}
The theoretical formulation and applicational exploitation of the
geoid in the context of general relativity has lately begun to attract
interest from geodesists and relativists alike \cite{kopeikin_towards_2015,oltean_geoids_2016,philipp_definition_2017}. 
Previously, the geoid
had only been well defined and worked with in (post-)Newtonian gravity \cite{hofmann-wellenhof_physical_2006,kopeikin_relativistic_2011}, wherein
it can essentially be thought of as a surface of constant potential, 
with applications ranging from the calibration of height measurements
of GPS satellites to the study of geophysical processes, climate patterns,
oceanic tides etc.

Lacking a notion of  ``potential'' in general relativity,
two possibilities\textemdash dubbed the \emph{a-geoid} and the \emph{u-geoid}\textemdash have
been variously propounded and discussed in the recent literature for
defining the geoid in a fully general-relativistic setting: first
in ref. \cite{kopeikin_towards_2015}, henceforth the \emph{KMK paper}
(after the authors); then in ref. \cite{oltean_geoids_2016}, henceforth
the \emph{GQF paper }(after the title); and most recently in ref.
\cite{philipp_definition_2017}, henceforth the \emph{ICS paper} (after
the title).

While a- and u- ``relativistic'' geoids had previously been defined (see, e.g., refs. \cite{kopeikin_1991,kopeikin_relativistic_2011}), until recently  analysis of such constructs had been essentially restricted to the low order post-Newtonian context, wherein they reduce to a surface of constant potential for some potential function (the Newtonian gravitational potential in the Newtonian limit). To our knowledge, refs. \cite{kopeikin_towards_2015,oltean_geoids_2016,philipp_definition_2017} constitute the first (\emph{non-approximate}) treatments of the geoid in the full theory of general relativity.

Our intent, first, is to clarify the relation between
these proposals, as well as the two (a- and u-) geoid definitions appearing
therein\textemdash the essential ideas of which we begin by summarizing
heuristically as follows.

\paragraph*{(1) The a-geoid:}

One demands that observers have zero acceleration tangential to the geoid; any acceleration is perpendicular to the geoid. (The Newtonian gravity analogue of 
this condition is that observers
lie on a surface of constant potential, 
which includes a centrifugal potential term in the case of a rotating geoid.)

\paragraph*{(2) The u-geoid:}

One demands that the clocks of observers on the geoid ``run at the
same speed''. (This condition has no Newtonian gravity analogue.)

These two definitions are stated side-by-side in general-relativistic
language already in the KMK paper \cite{kopeikin_towards_2015}. But crucially, there, the authors
impose from the beginning the restrictive assumption of spacetime
stationarity\textemdash under which, as indeed they point out, the
two definitions turn out to be mathematically equivalent. They then
proceed to derive, in this particular case, the PDEs that a geoid
should satisfy assuming a certain simplified model of the Earth.

In contrast to this,  we put forth in the GQF paper \cite{oltean_geoids_2016} a formulation of the geoid in
general relativity based on the first (a-geoid) idea, without imposing
\emph{any} assumptions on the spacetime. In fact, this formulation
emerges naturally as a particular choice (or ``gauge'') of a previously
developed and more general geometrical construction called a \emph{quasilocal
frame}\textemdash that is, essentially, a choice of a two-parameter
family of timelike worldlines comprising the worldtube boundary of
the history of a finite spatial volume. There are three degrees of freedom in the direction of the four-velocity vector tangent to each worldline, and so we can at least in principle\textemdash a point we shall elaborate upon below\textemdash impose three constraints on the motion of the quasilocal observers, thereby fixing
the specific nature of the quasilocal frame. In most of the past work
done on this 
\cite{epp_rigid_2009,mcgrath_quasilocal_2012,epp_existence_2012,epp_momentum_2013,mcgrath_post-newtonian_2014}, the
quasilocal frames were chosen to be rigid\textemdash called \emph{Rigid Quasilocal Frames} (RQFs), a natural choice when one is interested in fluxes of gravitational energy, momentum, and angular momentum across the quasilocal frame boundary. However, they can just as well
be chosen to describe geoids \cite{oltean_geoids_2016}. Thus, we
called them \emph{Geoid Quasilocal Frames} (GQFs). Moreover, we found
solutions for these GQFs in some spacetimes of interest, focusing on the (perturbed) Schwarzschild and Kerr metrics.

Recently, the ICS paper \cite{philipp_definition_2017} developed the idea of the u-geoid in the form
of a geometrical conception known as an \emph{isochronometric surface}
(ICS). These are surfaces that foliate a stationary spacetime according to
the requirement that one obtains along them a constant ``redshift
potential''\textemdash essentially, a measure of zero redshift between
any pair of observers on the ICS.
However, just as in the KMK paper, the assumption of stationarity
is imposed from the beginning\textemdash and is, in fact, necessary
for the very definition of an ICS in the first place. In fact, the main inspiration thereof was one of the earliest (more technologically-focused) attempts to formulate relativistic geodesy \cite{bjerhammar_relativistic_1986}, which explored the same u-geoid idea conceptually, albeit in significantly less mathematical profusion and with very restricted forms of the spacetime metric. Then the authors in ref. \cite{philipp_definition_2017}
recover many of the same solutions as we found for GQFs, up to rotation
terms (appearing if observers are attached to the surface of a ``rotating
Earth'' as opposed to a ``non-rotating'' one) which the GQF paper
did not consider.

%%%%%%%%%%%%%%%%%%%%%%%%%%%%%%%%%%%%%%
\vspace{5pt}
%%%%%%%%%%%%%%%%%%%%%%%%%%%%%%%%%%%%%%

\noindent \textit{Comparison of geoid definitions.} \textemdash{}
Now, we would like to put into perspective the scope of the overlap,
to the degree that it exists, between these approaches  insofar as their
mathematical formulations are concerned.

It is in fact not difficult to see that the geoid definitions in all
three papers are mathematically equivalent \emph{provided that the
spacetime is stationary} (and the geoid observers move along the associated timelike Killing vector field). We stress again however that, unlike the
GQF paper, the KMK and ICS papers impose this assumption from the
start (and formulate their definitions of the geoid accordingly);
and indeed, it is actually trivial to see that the defining (u-)geoid
equation in the KMK paper (their Eqn. (39) = constant) is identical
to that in the ICS paper (their Eqn. (17) = constant)---which, in turn, agrees with that given in ref. \cite{bjerhammar_relativistic_1986} (Eqn. (26) therein).

\begin{figure} \begin{center} \includegraphics[scale=0.34]{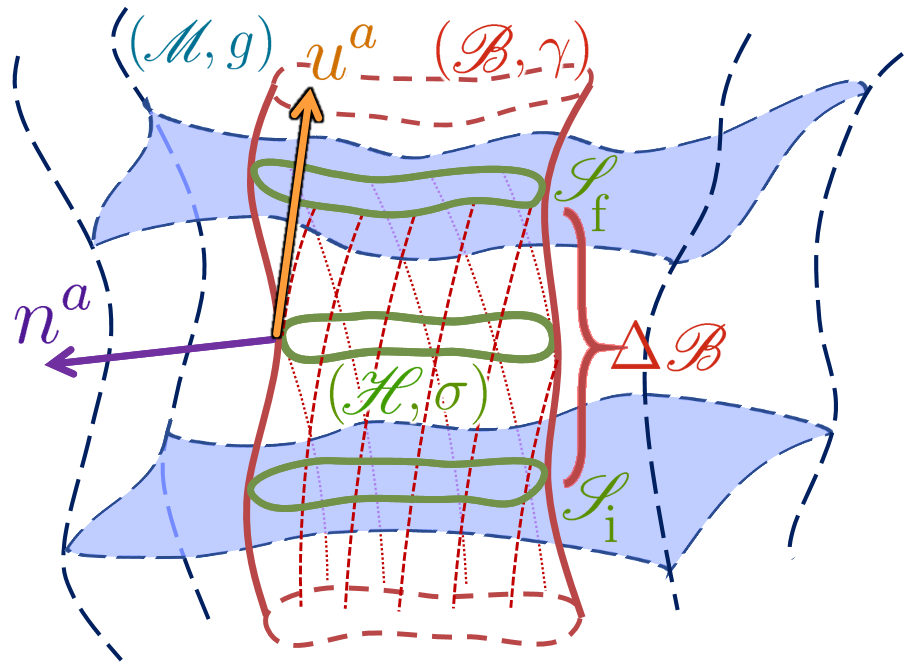} \caption{Definition of a quasilocal frame, illustrated for ease of visualization in the situation where the orthogonal subspaces $\mathscr{H}$ are closed two-surfaces.}\label{fig:qf} \end{center} \end{figure}

Notations differ, and so let us establish a common one (mirroring ours
in the GQF paper) in order to render the discussion here more precise.
Let the spacetime be $(\mathscr{M},\bm{g})$, with coordinates $\{x^{a}\}$.
In general, we use $\bm{T}$ to denote a $(k,l)$-tensor in $\mathscr{M}$
with abstract index notation $T^{a_{1}\cdots a_{k}}\,_{b_{1}\cdots b_{l}}$.
Let $u^{a}$ denote the timelike unit vector field tangent to the congruence of worldlines of the two-parameter family of observers comprising the geoid. This congruence is a submanifold of $\mathscr{M}$ (with topology $\mathbb{R}\times\mathbb{S}^{2}$) that we will call $\mathscr{B}$. Let $n^{a}$ be the outward-pointing
unit vector field normal to $\mathscr{B}$\textemdash which is uniquely fixed once $\bm{u}$ is
specified. Let $\mathscr{H}$
be the two-dimensional ``spatial'' subspace of the tangent space of $\mathscr{B}$ that is orthogonal
to $\bm{u}$. (A pictorial representation of this setup is given
in Figure \ref{fig:qf}, here taken from the GQF paper.) Let $\bm{\sigma}$ denote the two-dimensional ``spatial'' metric that projects tensor indices into $\mathscr{H}$, and is induced on $\mathscr{B}$ by the choice of $\bm{u}$ (and thus also $\bm{n}$):  $\sigma_{ab}=g_{ab}-n_{a}n_{b}+u_{a}u_{b}$. Let $\{x^{\mathfrak{i}}\}_{\mathfrak{i}=1}^{2}$ be ``spatial'' coordinates on  $\mathscr{B}$ that label the observers' worldlines, and let $t$ be a ``time'' coordinate on $\mathscr{B}$ such that surfaces of constant $t$ foliate $\mathscr{B}$ by spatial two-surfaces with topology $\mathbb{S}^{2}$. Finally, let $N$ denote the associated lapse function such that $\bm{u}=N^{-1}\partial/\partial t$. 

We can think of the (instantaneous) geoid as a surface of constant $t$ (as indicated in Figure \ref{fig:qf}), but note that \emph{in general} $\bm{u}$ need not be hypersurface orthogonal within $\mathscr{B}$ (this happens, e.g., when the geoid is rotating), in which case there exists no foliation of $\mathscr{B}$ by surfaces orthogonal to $\bm{u}$. In other words, the geoid observers are, \emph{in general}, \emph{in motion} with respect to the surfaces of constant $t$, and they will not be able to agree on a surface of simultaneity because the $\mathscr{H}$ spaces are not integrable. This is important, because we will often deal with projections of tensor components into the $\mathscr{H}$ spaces, which is not the same as projections tangential to a constant $t$ surface.

%--------------------- 
%---------------------

In this notation, let us temporarily restrict ourselves to stationary spacetimes, with the geoid observers moving along the timelike Killing vector field proportional to $\partial/\partial t$. Then, the defining (u-)geoid equation in the KMK
and ICS papers (their Eqns. (39) and (17), respectively) is in both
cases simply expressed in terms of the lapse function $N$
as
\begin{equation}
0=\partial_{\mathfrak{i}}N\,,\quad\textrm{for }\left(\mathscr{M},\bm{g}\right)\textrm{ stationary}\,.\label{eq:geoid_eqn_kmk_ics}
\end{equation}
The intuition behind this is that $N={\rm d}\tau/{\rm d}t$ is the
rate of change of proper time, and so (\ref{eq:geoid_eqn_kmk_ics})
is essentially saying that this should be the same for each of the geoid observers on a given surface of constant $t$ (clocks
are seen to ``run at the same speed''), and the same for all such surfaces of constant $t$ (since $\partial N/\partial t =0$ by the stationarity restriction). Moreover, it is not difficult
to follow the reasoning of both the KMK and ICS papers to the effect
that the a-geoid indeed coincides with the u-geoid under this restriction to stationarity. 

In contrast, in the GQF paper, we assume nothing about the spacetime
a priori. Our defining (a-)geoid equations (Eqn. (4.1) in ref. \cite{oltean_geoids_2016})
are completely general for \emph{any} spacetime $(\mathscr{M},\bm{g})$:
let $a^{a}=\nabla_{\bm{u}}u^{a}$ be the acceleration and $\theta_{ab}=\sigma_{a}^{\phantom{a}c}\sigma_{b}^{\phantom{b}d}\nabla_{c}u_{d}$
the strain rate tensor associated with the congruence; then what we
require of a geoid is simply the vanishing of the $\mathscr{H}$ component of the acceleration, $\alpha_{a}=\sigma_{a}^{\phantom{a}b}a_{b}$, as well as that
of the scalar expansion, $\theta=\sigma^{ab}\theta_{ab}$:
\begin{equation}
\begin{cases}
0=\alpha_{a}\,,\\
0=\theta\,.
\end{cases}\label{eq:geoid_defn_gqf1}
\end{equation}
In coordinate language, these can be equivalently stated, respectively, in the form of the
following \emph{time-dependent} system of PDEs (Eqns. (4.2)-(4.3)
in ref. \cite{oltean_geoids_2016}):
\begin{equation}
\begin{cases}
0=\partial_{\mathfrak{i}}N+\dot{u}_{\mathfrak{i}}\,,\\
0=\sigma^{\mathfrak{ij}}\dot{\sigma}_{\mathfrak{ij}}\,.
\end{cases}\label{eq:geoid_defn_gqf2}
\end{equation}

The first condition, which captures the essence of the heuristic a-geoid definition given in the Introduction, fixes \emph{two} of the three degrees of freedom of the geoid observers' motion.

Of course this is not sufficient for a sensible definition of a geoid; we must also essentially fix the normal component of the observers' acceleration, $\bm{n}\cdot\bm{a}$. Starting with a ``static'' geoid, by increasing/decreasing $\bm{n}\cdot\bm{a}$ we could make the geoid expand/contract arbitrarily. In the KMK and ICS papers this is not a consideration, since the spacetime is stationary, with observers moving along the timelike Killing vector field\textemdash trivially making the geoid static, and implicitly fixing $\bm{n}\cdot\bm{a}$. In a general, non-stationary spacetime, however, this issue must be explicitly considered, and the obvious, most geometrically natural choice is to demand zero scalar expansion of the congruence (which still allows for a dynamic shear, necessary to carry gravitational radiation through the geoid\textemdash more on this below). Then $\bm{n}\cdot\bm{a}$ is implicitly fixed through a Raychaudhuri-like equation.
This condition on the scalar expansion ($\theta=0$) fixes an additional
\emph{one} degree of freedom. Together with $\alpha_{a}=0$, this fixes 
\emph{three} degrees of freedom, completely
specifying the quasilocal frame, i.e., in essence completely determining the three degrees of freedom in specifying the vector field $\bm{u}$
that defines the congruence.

It is obvious by inspection that if a stationary spacetime is assumed (with the geoid observers moving along the timelike Killing vector field),
the time-dependent terms in (\ref{eq:geoid_defn_gqf2}) vanish; hence,
the second equation becomes trivial and the first simply reduces to
(\ref{eq:geoid_eqn_kmk_ics}), that is, the geoid definition of the
KMK and ICS papers.

%%%%%%%%%%%%%%%%%%%%%%%%%%%%%%%%%%%%%%
\vspace{5pt}
%%%%%%%%%%%%%%%%%%%%%%%%%%%%%%%%%%%%%%

\noindent \textit{Time dependence.} \textemdash{} We wish now to examine
in greater depth the role of time in defining the relativistic geoid.

The contention of the ICS paper is that theirs is a ``more operational''
formulation since in order to prescribe the geoid, one simply needs
(according to them) ``precise clocks'' that ``run at the same speed''.
This is equivalent to saying that there should be zero redshift between \emph{any
two} observers on the geoid. In particular,
then, let us apply this statement to \emph{nearest-neighbor} pairs
of observers, and to keep the analysis completely general, let us consider a general quasilocal frame (one with \emph{no} conditions, such as the GQF or RQF conditions, imposed). Let $k^{a}$ be the tangent to an affinely-parametrized null geodesic $\lambda$
along which a light ray is assumed to travel, and let $\gamma$ be
the worldline of an observer in the congruence with four-velocity $\bm{u}$.
Then the nearest-neighbor zero redshift requirement is simply $0=\nabla_{\bm{k}}(\bm{u}\cdot\bm{k})$.

Let us analyze what this is saying. Since $\bm{k}$ is (necessarily)
tangent to $\mathscr{B}$ (nearest-neighbors), we can decompose it into $k^{a}|_{\gamma}=u^{a}+N^{a}$
where $\bm{N}$ is a unit vector tangent to $\mathscr{H}$ defining the direction of the light ray towards a nearest neighbor (i.e., one angular degree of freedom). Using this, along with the fact that $\nabla_{\bm{k}}k^{a}=0$
along $\lambda$, a straightforward calculation reveals:
\begin{align}
\nabla_{\bm{k}}(\bm{u}\cdot\bm{k})|_{\gamma} & =k^{a}k^{b}\nabla_{a}u_{b}\label{eq:redshift_formula1}\\
 & =\frac{1}{2}\theta+\alpha_{a}N^{a}+\theta_{\langle ab\rangle}N^{a}N^{b}\,,\label{eq:redshift_formula2}
\end{align}
where $\theta_{\langle ab\rangle}$ is the symmetric trace-free (STF)
or ``shear'' part of $\bm{\theta}$, which has two independent components. The three terms in the last line
can be regarded as the $\ell=0,1,2$ spherical harmonic terms (respectively),
and so the geoid definition of the ICS paper (i.e. the vanishing of
(\ref{eq:redshift_formula1}) for all $N^{a}$, to ensure zero redshift)
holds if and only if, one by one, we have $\theta=0$, $\alpha_{a}=0$
and $\theta_{\langle ab\rangle}=0$. These are \emph{five} constraints
in total.

Now, the three constraints $\theta=0=\alpha_{a}$ (without $\theta_{\langle ab\rangle}=0$)
constitute precisely the geoid definition (\ref{eq:geoid_defn_gqf1})
from our GQF paper. Moreover, the three constraints $\theta=0=\theta_{\langle ab\rangle}\Leftrightarrow0=\theta_{(ab)}$
(without $\alpha_{a}=0$) are actually the equations previously employed
to define \emph{rigid} quasilocal frames (RQFs) \cite{epp_rigid_2009,mcgrath_quasilocal_2012,epp_existence_2012,epp_momentum_2013,mcgrath_post-newtonian_2014}.
Either set of three constraints is logically consistent with the availability
of the (three) degrees of freedom in the motion of observers in an arbitrary spacetime.

However, imposing all \emph{five} constraints $\alpha^{a}=0=\theta_{(ab)}$
\emph{over-determines} the available degrees of freedom. In other
words, they not only define the geoid (which only requires three of
these), but they inherently restrict (as a consequence of the additional
two) the geometry of the spacetime within which it can be embedded. In particular,
this means that the spacetime \emph{must} be stationary\textemdash as
recognized in the ICS paper itself. The authors' main point of argumentation
for doing this is that a ``more operational'' determination of the
geoid can thus be achieved via the use of standard clocks (connected
by optical fibers). Yet, this forces the very restrictive presupposition
that the geoid will not undergo any time evolution\textemdash for
if it does, then it is clear that it will not be possible to consistently
capture the details of this evolution just by using ICSs.

The two (u- and a-) geoid notions are thus, in general, \emph{not
}equivalent in general relativity. In other words, because of the
time-dependent terms in (\ref{eq:geoid_defn_gqf2}), one cannot\emph{\textemdash }except
in very special (stationary) situations\emph{\textemdash }equate surfaces
composed of clocks that ``run at the same speed'' (u-geoids) with
those relative to which observers accelerate only perpendicularly
(a-geoids). Observers on an a-geoid/GQF, determined (only) by $\theta=0=\alpha_{a}$,
\emph{will }in general, according to (\ref{eq:redshift_formula2}),
observe a \emph{time-dependent redshift }due to the shear (STF) part
of the strain rate tensor. This $\ell=2$ degree of freedom represents a ``true'' (general-relativistic) gravitational degree of freedom, and plays a crucial role vis-a-vis the gravitational radiation passing through the geoid. In 
ref.~\cite{mcgrath_quasilocal_2012} (see Eqn. (32) and the discussion following it), we identify, in the case of an RQF, $\alpha_a {\cal P}^a$ as the operationally defined gravitational power density for gravitational radiation passing through the RQF (where ${\cal P}^a$ is the quasilocal momentum density, measured with gyroscopes); similarly, in the case of a GQF, $\theta_{\langle ab \rangle}{\cal S}^{\langle ab \rangle}$ is the operationally defined gravitational power density for gravitational radiation passing through the GQF (where ${\cal S}^{\langle ab \rangle}$ is the STF part of the quasilocal stress tensor\textemdash two independent components); see ref. \cite{oltean_geoids_2016} for details.

We emphasize that a GQF is just as ``operationally'' defined as an ICS. By measuring nearest-neighbor redshifts, we can, according to (\ref{eq:redshift_formula2}), measure each of $\theta$, $\alpha_a$, and $\theta_{\langle ab \rangle}$\textemdash which can, of course, be measured by other means as well. Measuring these, and responding with the appropriate normal acceleration, as necessary, observers can maintain a GQF frame, and even measure the flow of gravitational energy, momentum, and angular momentum through that frame. See refs. \cite{epp_rigid_2009,mcgrath_quasilocal_2012,epp_existence_2012,epp_momentum_2013,mcgrath_post-newtonian_2014} for more details in the case of RQFs, and ref. \cite{oltean_geoids_2016} for the case of GQFs.

%%%%%%%%%%%%%%%%%%%%%%%%%%%%%%%%%%%%%%
\vspace{5pt}
%%%%%%%%%%%%%%%%%%%%%%%%%%%%%%%%%%%%%%

\noindent \textit{Existence of GQFs.} \textemdash{} We have generally
claimed so far that GQFs can be constructed in any arbitrary spacetime.
However, the precise question of existence of any sort of (``gauge''
fixed) quasilocal frame (be it a GQF, or RQF, or anything else) is
in fact much more subtle and elaborate than perhaps we have implied
so far, and certainly than can be explored adequately in this note;
nevertheless, this point warrants here a bit more qualification.

We have seen that, intuitively, the three degrees of freedom in specifying
our quasilocal (two-parameter) family of observers correspond to the
freedom of imposing three constraints thereon\textemdash achieving
a GQF, an RQF etc. Yet, the mathematical problem of the existence
of solutions to these (three) equations must be dealt with in each
separate case. In other words, in order to guarantee that ``GQFs always
exist'', for instance, it is actually necessary to prove the following:
that for any arbitrary spacetime $(\mathscr{M},\bm{g})$, one can
always construct a GQF\textemdash that is to say, \emph{solutions
to the GQF equations (\ref{eq:geoid_defn_gqf1}) exist}\textemdash 
on an appropriately fibered timelike worldtube ``centered'' around
any desired timelike worldline $\Gamma$ in $\mathscr{M}$. 

This is a nontrivial problem. Existence of RQFs was shown in ref.
\cite{epp_existence_2012}; the strategy was to construct, using a
Fermi normal coordinates approach, the general solution to the RQF
equations as a perturbative series in powers of the areal radius (about
the trivial, small-sphere RQF), and to show that there is in principle no technical
obstruction to extending these solutions to any desired order. 

Following an analogous approach, it is also possible to prove this
for GQFs. We will present the full analysis in a forthcoming paper \cite{oltean_geoids_2017-future}.
This will essentially show that one can generically solve for \emph{arbitrary}
(time-dependent) GQFs in any spacetime.

%%%%%%%%%%%%%%%%%%%%%%%%%%%%%%%%%%%%%%
\vspace{5pt}
%%%%%%%%%%%%%%%%%%%%%%%%%%%%%%%%%%%%%%

\noindent \textit{Conclusions.} \textemdash{} Defining the relativistic
geoid as a u-geoid/ICS suffers from evident and unnecessary deficiencies
in comparison with the completely general a-geoid/GQF definition, from both
a foundational as well as pragmatic point of view.

Foundationally, it is highly unsatisfactory to have to impose a priori
such a restrictive assumption\textemdash stationarity\textemdash on
the spacetime itself in order to even be able to make sense the u-geoid/ICS,
especially when the a-geoid/GQF trivially reduces to it thereunder.
Indeed, the generic success and usefulness of the broader formalism
of quasilocal frames has lain precisely in eliminating the all-too-pervasive
need for stipulating special spacetimes (i.e. necessitating the existence of
Killing vector fields) in making sense of the mechanisms behind energy, momentum, and angular momentum transfer (conservation laws) in general relativity \cite{epp_rigid_2009,mcgrath_quasilocal_2012,epp_existence_2012,epp_momentum_2013,mcgrath_post-newtonian_2014};
in this regard, it is no different\textemdash as will be developed,
we expect, in much greater depth in the future\textemdash with geodetic
modeling.

Pragmatically, one of the principal motivations for moving geodesy
from the setting of Newtonian gravity to general relativity is increased
accuracy\textemdash and, as such, it is unclear how a geoid definition
based on stationarity can adequately handle the treatment of relevant
\emph{time-dependent} perturbations. With the uncertainty of the latest
generation of optical atomic clocks approaching the level of one part
in $10^{18}$, time variations in the geoid\textemdash corresponding
to an accuracy level of mm to cm\textemdash will have to be taken
into account \cite{voigt_time-variable_2016,muller_high_2017}. These
arise, for example, from relative perturbations of the solid Earth
tides, periodic effects due to ocean tides, non-tidal oceanic and
atmospheric effects, variations due to land hydrology, tectonic processes
etc.

The ICS paper suggests dealing with these by thinking of the ``true''
metric $\bm{g}$ as a sum of a stationary part $\bm{g}_{{\rm stat}}$
(which can include the time-averaged effects of those perturbations
that are cyclic), and a ``small'' time-dependent part $\bm{h}$, i.e.
$\bm{g}=\bm{g}_{{\rm stat}}+\bm{h}$. However, working out a ``perturbative
theory'' of the geoid (for $\bm{h}$) in this setting seems unnecessarily
contrived and potentially confusing.
One would need a criterion for constructing the the components of the ``stationary'' part of the metric $\bm{g}_{{\rm stat}}$ in a time-dependent setting where, by definition, the ICS's change with time. It is not at
all apparent how one could deduce  related changes in
$\bm{h}$, without having (from the beginning) a theory of time-dependent ICS's.

Of course, this is not at all an issue if one uses the a-geoid/GQF
approach to define the geoid. Being applicable in arbitrary spacetimes,
the inclusion and computation of time-dependent perturbations fits
naturally into our general formalism\textemdash as, indeed, we will
illustrate mathematically in our future paper on GQF existence
\cite{oltean_geoids_2017-future}. 
Arbitrarily
large time variations in strong gravitational fields\textemdash unnecessary
for cases such as the geoid of the Earth, but potentially relevant
for astrophysical applications\textemdash can certainly also be accommodated
with GQFs.

As mentioned earlier, and emphasized again in closing, a time-dependent problem of particular interest is that of gravitational
waves passing through the geoid; the effect thereof is felt in the
form of shearing, which can be measured via the $\ell = 2$ component of the time-dependent nearest-neighbor redshift.
The gravitational energy flux through the geoid is\textemdash as we have mentioned, and as discussed
at greater length in the GQF paper\textemdash directly related
to this quantity (contracted with the STF part of the quasilocal stress). In our forthcoming paper, we will additionally offer
a more detailed discussion on gravitational waves in the context of
GQFs.

\section*{Acknowledgements}

This work was supported in part by the Natural Sciences and Engineering Research Council of Canada.

%%%%%%%%%%%%%%%%%%%%%%%%%%%%%%%%%%%%%%
\bibliography{references}

\begin{thebibliography}{15}%
\makeatletter
\providecommand \@ifxundefined [1]{%
 \@ifx{#1\undefined}
}%
\providecommand \@ifnum [1]{%
 \ifnum #1\expandafter \@firstoftwo
 \else \expandafter \@secondoftwo
 \fi
}%
\providecommand \@ifx [1]{%
 \ifx #1\expandafter \@firstoftwo
 \else \expandafter \@secondoftwo
 \fi
}%
\providecommand \natexlab [1]{#1}%
\providecommand \enquote  [1]{``#1''}%
\providecommand \bibnamefont  [1]{#1}%
\providecommand \bibfnamefont [1]{#1}%
\providecommand \citenamefont [1]{#1}%
\providecommand \href@noop [0]{\@secondoftwo}%
\providecommand \href [0]{\begingroup \@sanitize@url \@href}%
\providecommand \@href[1]{\@@startlink{#1}\@@href}%
\providecommand \@@href[1]{\endgroup#1\@@endlink}%
\providecommand \@sanitize@url [0]{\catcode `\\12\catcode `\$12\catcode
  `\&12\catcode `\#12\catcode `\^12\catcode `\_12\catcode `\%12\relax}%
\providecommand \@@startlink[1]{}%
\providecommand \@@endlink[0]{}%
\providecommand \url  [0]{\begingroup\@sanitize@url \@url }%
\providecommand \@url [1]{\endgroup\@href {#1}{\urlprefix }}%
\providecommand \urlprefix  [0]{URL }%
\providecommand \Eprint [0]{\href }%
\providecommand \doibase [0]{http://dx.doi.org/}%
\providecommand \selectlanguage [0]{\@gobble}%
\providecommand \bibinfo  [0]{\@secondoftwo}%
\providecommand \bibfield  [0]{\@secondoftwo}%
\providecommand \translation [1]{[#1]}%
\providecommand \BibitemOpen [0]{}%
\providecommand \bibitemStop [0]{}%
\providecommand \bibitemNoStop [0]{.\EOS\space}%
\providecommand \EOS [0]{\spacefactor3000\relax}%
\providecommand \BibitemShut  [1]{\csname bibitem#1\endcsname}%
\let\auto@bib@innerbib\@empty
%</preamble>
\bibitem [{\citenamefont {Philipp}\ \emph {et~al.}(2017)\citenamefont
  {Philipp}, \citenamefont {Perlick}, \citenamefont {Puetzfeld}, \citenamefont
  {Hackmann},\ and\ \citenamefont {Lämmerzahl}}]{philipp_definition_2017}%
  \BibitemOpen
  \bibfield  {author} {\bibinfo {author} {\bibfnamefont {Dennis}\ \bibnamefont
  {Philipp}}, \bibinfo {author} {\bibfnamefont {Volker}\ \bibnamefont
  {Perlick}}, \bibinfo {author} {\bibfnamefont {Dirk}\ \bibnamefont
  {Puetzfeld}}, \bibinfo {author} {\bibfnamefont {Eva}\ \bibnamefont
  {Hackmann}}, \ and\ \bibinfo {author} {\bibfnamefont {Claus}\ \bibnamefont
  {Lämmerzahl}},\ }\bibfield  {title} {\enquote {\bibinfo {title} {Definition
  of the relativistic geoid in terms of isochronometric surfaces},}\ }\href
  {\doibase 10.1103/PhysRevD.95.104037} {\bibfield  {journal} {\bibinfo
  {journal} {Physical Review D}\ }\textbf {\bibinfo {volume} {95}},\ \bibinfo
  {pages} {104037} (\bibinfo {year} {2017})}\BibitemShut {NoStop}%
\bibitem [{\citenamefont {Oltean}\ \emph {et~al.}(2016)\citenamefont {Oltean},
  \citenamefont {Epp}, \citenamefont {McGrath},\ and\ \citenamefont
  {Mann}}]{oltean_geoids_2016}%
  \BibitemOpen
  \bibfield  {author} {\bibinfo {author} {\bibfnamefont {Marius}\ \bibnamefont
  {Oltean}}, \bibinfo {author} {\bibfnamefont {Richard~J.}\ \bibnamefont
  {Epp}}, \bibinfo {author} {\bibfnamefont {Paul~L.}\ \bibnamefont {McGrath}},
  \ and\ \bibinfo {author} {\bibfnamefont {Robert~B.}\ \bibnamefont {Mann}},\
  }\bibfield  {title} {\enquote {\bibinfo {title} {Geoids in general
  relativity: geoid quasilocal frames},}\ }\href {\doibase
  10.1088/0264-9381/33/10/105001} {\bibfield  {journal} {\bibinfo  {journal}
  {Classical and Quantum Gravity}\ }\textbf {\bibinfo {volume} {33}},\ \bibinfo
  {pages} {105001} (\bibinfo {year} {2016})}\BibitemShut {NoStop}%
\bibitem [{\citenamefont {Kopeikin}\ \emph {et~al.}(2015)\citenamefont
  {Kopeikin}, \citenamefont {Mazurova},\ and\ \citenamefont
  {Karpik}}]{kopeikin_towards_2015}%
  \BibitemOpen
  \bibfield  {author} {\bibinfo {author} {\bibfnamefont {Sergei~M.}\
  \bibnamefont {Kopeikin}}, \bibinfo {author} {\bibfnamefont {Elena~M.}\
  \bibnamefont {Mazurova}}, \ and\ \bibinfo {author} {\bibfnamefont
  {Alexander~P.}\ \bibnamefont {Karpik}},\ }\bibfield  {title} {\enquote
  {\bibinfo {title} {Towards an exact relativistic theory of {Earth}'s geoid
  undulation},}\ }\href {\doibase 10.1016/j.physleta.2015.02.046} {\bibfield
  {journal} {\bibinfo  {journal} {Physics Letters A}\ }\textbf {\bibinfo
  {volume} {379}},\ \bibinfo {pages} {1555--1562} (\bibinfo {year}
  {2015})}\BibitemShut {NoStop}%
\bibitem [{\citenamefont {Hofmann-Wellenhof}\ and\ \citenamefont
  {Moritz}(2006)}]{hofmann-wellenhof_physical_2006}%
  \BibitemOpen
  \bibfield  {author} {\bibinfo {author} {\bibfnamefont {Bernhard}\
  \bibnamefont {Hofmann-Wellenhof}}\ and\ \bibinfo {author} {\bibfnamefont
  {Helmut}\ \bibnamefont {Moritz}},\ }\href@noop {} {\emph {\bibinfo {title}
  {Physical {Geodesy}}}},\ \bibinfo {edition} {2nd}\ ed.\ (\bibinfo
  {publisher} {Springer},\ \bibinfo {address} {Wien ; New York},\ \bibinfo
  {year} {2006})\BibitemShut {NoStop}%
\bibitem [{\citenamefont {Kopeikin}\ \emph {et~al.}(2011)\citenamefont
  {Kopeikin}, \citenamefont {Efroimsky},\ and\ \citenamefont
  {Kaplan}}]{kopeikin_relativistic_2011}%
  \BibitemOpen
  \bibfield  {author} {\bibinfo {author} {\bibfnamefont {Sergei}\ \bibnamefont
  {Kopeikin}}, \bibinfo {author} {\bibfnamefont {Michael}\ \bibnamefont
  {Efroimsky}}, \ and\ \bibinfo {author} {\bibfnamefont {George}\ \bibnamefont
  {Kaplan}},\ }\href@noop {} {\emph {\bibinfo {title} {Relativistic {Celestial}
  {Mechanics} of the {Solar} {System}}}}\ (\bibinfo  {publisher} {Wiley-VCH},\
  \bibinfo {address} {Weinheim},\ \bibinfo {year} {2011})\BibitemShut {NoStop}%
\bibitem [{\citenamefont {{Kopeikin}}(1991)}]{kopeikin_1991}%
  \BibitemOpen
  \bibfield  {author} {\bibinfo {author} {\bibfnamefont {S.~M.}\ \bibnamefont
  {{Kopeikin}}},\ }\bibfield  {title} {\enquote {\bibinfo {title}
  {{Relativistic Manifestations of gravitational fields in gravimetry and
  geodesy}},}\ }\href@noop {} {\bibfield  {journal} {\bibinfo  {journal}
  {Manuscripta Geodaetica}\ }\textbf {\bibinfo {volume} {16}} (\bibinfo {year}
  {1991})}\BibitemShut {NoStop}%
\bibitem [{\citenamefont {Epp}\ \emph {et~al.}(2009)\citenamefont {Epp},
  \citenamefont {Mann},\ and\ \citenamefont {McGrath}}]{epp_rigid_2009}%
  \BibitemOpen
  \bibfield  {author} {\bibinfo {author} {\bibfnamefont {Richard~J.}\
  \bibnamefont {Epp}}, \bibinfo {author} {\bibfnamefont {Robert~B.}\
  \bibnamefont {Mann}}, \ and\ \bibinfo {author} {\bibfnamefont {Paul~L.}\
  \bibnamefont {McGrath}},\ }\bibfield  {title} {\enquote {\bibinfo {title}
  {Rigid motion revisited: rigid quasilocal frames},}\ }\href {\doibase
  10.1088/0264-9381/26/3/035015} {\bibfield  {journal} {\bibinfo  {journal}
  {Classical and Quantum Gravity}\ }\textbf {\bibinfo {volume} {26}},\ \bibinfo
  {pages} {035015} (\bibinfo {year} {2009})}\BibitemShut {NoStop}%
\bibitem [{\citenamefont {McGrath}\ \emph {et~al.}(2012)\citenamefont
  {McGrath}, \citenamefont {Epp},\ and\ \citenamefont
  {Mann}}]{mcgrath_quasilocal_2012}%
  \BibitemOpen
  \bibfield  {author} {\bibinfo {author} {\bibfnamefont {Paul~L.}\ \bibnamefont
  {McGrath}}, \bibinfo {author} {\bibfnamefont {Richard~J.}\ \bibnamefont
  {Epp}}, \ and\ \bibinfo {author} {\bibfnamefont {Robert~B.}\ \bibnamefont
  {Mann}},\ }\bibfield  {title} {\enquote {\bibinfo {title} {Quasilocal
  conservation laws: why we need them},}\ }\href {\doibase
  10.1088/0264-9381/29/21/215012} {\bibfield  {journal} {\bibinfo  {journal}
  {Classical and Quantum Gravity}\ }\textbf {\bibinfo {volume} {29}},\ \bibinfo
  {pages} {215012} (\bibinfo {year} {2012})}\BibitemShut {NoStop}%
\bibitem [{\citenamefont {Epp}\ \emph {et~al.}(2012)\citenamefont {Epp},
  \citenamefont {Mann},\ and\ \citenamefont {McGrath}}]{epp_existence_2012}%
  \BibitemOpen
  \bibfield  {author} {\bibinfo {author} {\bibfnamefont {Richard~J.}\
  \bibnamefont {Epp}}, \bibinfo {author} {\bibfnamefont {Robert~B.}\
  \bibnamefont {Mann}}, \ and\ \bibinfo {author} {\bibfnamefont {Paul~L.}\
  \bibnamefont {McGrath}},\ }\bibfield  {title} {\enquote {\bibinfo {title} {On
  the {Existence} and {Utility} of {Rigid} {Quasilocal} {Frames}},}\ }in\
  \href@noop {} {\emph {\bibinfo {booktitle} {Classical and {Quantum}
  {Gravity}: {Theory}, {Analysis} and {Applications}}}},\ \bibinfo {series and
  number} {Physics {Research} and {Technology}},\ \bibinfo {editor} {edited by\
  \bibinfo {editor} {\bibfnamefont {Vincent~R.}\ \bibnamefont {Frignanni}}}\
  (\bibinfo  {publisher} {Nova Science Publishers},\ \bibinfo {address} {New
  York},\ \bibinfo {year} {2012})\BibitemShut {NoStop}%
\bibitem [{\citenamefont {Epp}\ \emph {et~al.}(2013)\citenamefont {Epp},
  \citenamefont {McGrath},\ and\ \citenamefont {Mann}}]{epp_momentum_2013}%
  \BibitemOpen
  \bibfield  {author} {\bibinfo {author} {\bibfnamefont {Richard~J.}\
  \bibnamefont {Epp}}, \bibinfo {author} {\bibfnamefont {Paul~L.}\ \bibnamefont
  {McGrath}}, \ and\ \bibinfo {author} {\bibfnamefont {Robert~B.}\ \bibnamefont
  {Mann}},\ }\bibfield  {title} {\enquote {\bibinfo {title} {Momentum in
  general relativity: local versus quasilocal conservation laws},}\ }\href
  {\doibase 10.1088/0264-9381/30/19/195019} {\bibfield  {journal} {\bibinfo
  {journal} {Classical and Quantum Gravity}\ }\textbf {\bibinfo {volume}
  {30}},\ \bibinfo {pages} {195019} (\bibinfo {year} {2013})}\BibitemShut
  {NoStop}%
\bibitem [{\citenamefont {McGrath}\ \emph {et~al.}(2014)\citenamefont
  {McGrath}, \citenamefont {Chanona}, \citenamefont {Epp}, \citenamefont
  {Koop},\ and\ \citenamefont {Mann}}]{mcgrath_post-newtonian_2014}%
  \BibitemOpen
  \bibfield  {author} {\bibinfo {author} {\bibfnamefont {Paul~L.}\ \bibnamefont
  {McGrath}}, \bibinfo {author} {\bibfnamefont {Melanie}\ \bibnamefont
  {Chanona}}, \bibinfo {author} {\bibfnamefont {Richard~J.}\ \bibnamefont
  {Epp}}, \bibinfo {author} {\bibfnamefont {Michael~J.}\ \bibnamefont {Koop}},
  \ and\ \bibinfo {author} {\bibfnamefont {Robert~B.}\ \bibnamefont {Mann}},\
  }\bibfield  {title} {\enquote {\bibinfo {title} {Post-{Newtonian}
  conservation laws in rigid quasilocal frames},}\ }\href {\doibase
  10.1088/0264-9381/31/9/095006} {\bibfield  {journal} {\bibinfo  {journal}
  {Classical and Quantum Gravity}\ }\textbf {\bibinfo {volume} {31}},\ \bibinfo
  {pages} {095006} (\bibinfo {year} {2014})}\BibitemShut {NoStop}%
\bibitem [{\citenamefont {Bjerhammar}(1986)}]{bjerhammar_relativistic_1986}%
  \BibitemOpen
  \bibfield  {author} {\bibinfo {author} {\bibfnamefont {Arne}\ \bibnamefont
  {Bjerhammar}},\ }\href
  {https://www.ngs.noaa.gov/PUBS_LIB/RelativisticGeodesy_TR_NOS118_NGS36.pdf}
  {\emph {\bibinfo {title} {Relativistic {Geodesy}}}},\ \bibinfo {type}
  {Technical {Report}}\ \bibinfo {number} {NOS 118 NGS 36}\ (\bibinfo
  {institution} {NOAA},\ \bibinfo {address} {Rockville, MD},\ \bibinfo {year}
  {1986})\BibitemShut {NoStop}%
\bibitem [{\citenamefont {Oltean}\ \emph {et~al.}()\citenamefont {Oltean},
  \citenamefont {Epp}, \citenamefont {McGrath},\ and\ \citenamefont
  {Mann}}]{oltean_geoids_2017-future}%
  \BibitemOpen
  \bibfield  {author} {\bibinfo {author} {\bibfnamefont {Marius}\ \bibnamefont
  {Oltean}}, \bibinfo {author} {\bibfnamefont {Richard~J.}\ \bibnamefont
  {Epp}}, \bibinfo {author} {\bibfnamefont {Paul~L.}\ \bibnamefont {McGrath}},
  \ and\ \bibinfo {author} {\bibfnamefont {Robert~B.}\ \bibnamefont {Mann}},\
  }\bibfield  {title} {\enquote {\bibinfo {title} {Work in progress},}\
  }\href@noop {} {\ }\BibitemShut {NoStop}%
\bibitem [{\citenamefont {Voigt}\ \emph {et~al.}(2016)\citenamefont {Voigt},
  \citenamefont {Denker},\ and\ \citenamefont
  {Timmen}}]{voigt_time-variable_2016}%
  \BibitemOpen
  \bibfield  {author} {\bibinfo {author} {\bibfnamefont {C.}~\bibnamefont
  {Voigt}}, \bibinfo {author} {\bibfnamefont {H.}~\bibnamefont {Denker}}, \
  and\ \bibinfo {author} {\bibfnamefont {L.}~\bibnamefont {Timmen}},\
  }\bibfield  {title} {\enquote {\bibinfo {title} {Time-variable gravity
  potential components for optical clock comparisons and the definition of
  international time scales},}\ }\href {\doibase 10.1088/0026-1394/53/6/1365}
  {\bibfield  {journal} {\bibinfo  {journal} {Metrologia}\ }\textbf {\bibinfo
  {volume} {53}},\ \bibinfo {pages} {1365} (\bibinfo {year}
  {2016})}\BibitemShut {NoStop}%
\bibitem [{\citenamefont {Müller}\ \emph {et~al.}(2017)\citenamefont
  {Müller}, \citenamefont {Dirkx}, \citenamefont {Kopeikin}, \citenamefont
  {Lion}, \citenamefont {Panet}, \citenamefont {Petit},\ and\ \citenamefont
  {Visser}}]{muller_high_2017}%
  \BibitemOpen
  \bibfield  {author} {\bibinfo {author} {\bibfnamefont {J.}~\bibnamefont
  {Müller}}, \bibinfo {author} {\bibfnamefont {D.}~\bibnamefont {Dirkx}},
  \bibinfo {author} {\bibfnamefont {S.~M.}\ \bibnamefont {Kopeikin}}, \bibinfo
  {author} {\bibfnamefont {G.}~\bibnamefont {Lion}}, \bibinfo {author}
  {\bibfnamefont {I.}~\bibnamefont {Panet}}, \bibinfo {author} {\bibfnamefont
  {G.}~\bibnamefont {Petit}}, \ and\ \bibinfo {author} {\bibfnamefont {P.~N.
  A.~M.}\ \bibnamefont {Visser}},\ }\bibfield  {title} {\enquote {\bibinfo
  {title} {High {Performance} {Clocks} and {Gravity} {Field}
  {Determination}},}\ }\href {http://arxiv.org/abs/1702.06761} {\bibfield
  {journal} {\bibinfo  {journal} {arXiv:1702.06761 [physics.geo-ph]}\ }
  (\bibinfo {year} {2017})}\BibitemShut {NoStop}%
\end{thebibliography}%
%%%%%%%%%%%%%%%%%%%%%%%%%%%%%%%%%%%%%%

\end{document}